\documentclass{aa}  
\usepackage{graphicx}
\usepackage{txfonts,textcomp}
\usepackage{amsmath}   
\usepackage{gensymb}
\usepackage[normalem]{ulem}
\usepackage{array}
\usepackage{float}
\usepackage{lipsum}
\usepackage{adjustbox}

\usepackage{threeparttable}
\newcommand\tstrut{\rule{0pt}{2.4ex}}

\newcommand{\src}{{\it J1727}} 
\newcommand{\srclong}{{\it Swift J1727.8--1613}}

\usepackage[colorlinks =TRUE, linkcolor = blue,
            urlcolor  = blue,
            citecolor = blue,
            anchorcolor = blue]{hyperref}
\begin{document} 

\title{Detection of a type-C quasi-periodic oscillation during the soft-to-hard transition in Swift J1727.8 --1613}
   \author{\href{https://orcid.org/0009-0004-1197-5935}{Maïmouna Brigitte\thanks{E-mail:maimouna.brigitte@asu.cas.cz}\includegraphics[scale=0.7]{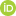}}\inst{1,}\inst{2} \and \href{https://orcid.org/0000-0002-5870-0443}
   {Noel Castro~Segura\includegraphics[scale=0.7]{ORCID-iD_icon_16x16.png}}\inst{3} \and \href{https://orcid.org/0000-0001-9072-4069}
   {Federico García\includegraphics[scale=0.7]{ORCID-iD_icon_16x16.png}}\inst{4,}\inst{5} \and \href{https://orcid.org/0000-0003-2931-0742}
   {Jiří Svoboda\includegraphics[scale=0.7]{ORCID-iD_icon_16x16.png}}\inst{1} \and \href{https://orcid.org/0000-0001-7796-4279}
   {María D\'iaz Trigo\includegraphics[scale=0.7]{ORCID-iD_icon_16x16.png}}\inst{6} \and \href{https://orcid.org/0000-0003-2187-2708}
   {Mariano M\'endez\includegraphics[scale=0.7]{ORCID-iD_icon_16x16.png}}\inst{7} \and 
   \href{https://orcid.org/0000-0002-1481-1870}
   {Federico M. Vincentelli \includegraphics[scale=0.7]{ORCID-iD_icon_16x16.png}}\inst{8,9} \and 
   \href{https://orcid.org/0000-0002-5341-6929}
   {Douglas J. K. Buisson \includegraphics[scale=0.7]{ORCID-iD_icon_16x16.png}}\inst{9,} \inst{10} 
   \and
   \href{https://orcid.org/0000-0002-3422-0074}
   {Diego Altamirano \includegraphics[scale=0.7]{ORCID-iD_icon_16x16.png}}\inst{11} 
   }
   \institute{
   Astronomical Institute of the Czech Academy of Sciences, Bo\v{c}n\'{i} II 1401/1, 14100 Prague 4, Czech Republic
   \and
   Astronomical Institute, Faculty of Mathematics and Physics, Charles University, V Holešovičkách 2, Prague 8, 18000, Czech Republic
   \and
   Department of Physics, University of Warwick, Gibbet Hill Road, Coventry CV4 7AL, UK
   \and
   Instituto Argentino de Radioastronomía (CCT La Plata, CONICET; CICPBA; UNLP), C.C.5, (1894) Villa Elisa, Buenos Aires, Argentina
   \and
   Facultad de Ciencias Astronómicas y Geofísicas, Universidad Nacional de La Plata, Paseo del Bosque, B1900FWA La Plata, Argentina
   \and
   ESO, Karl-Schwarzschild-Strasse 2, 85748, Garching bei München, Germany
   \and 
   Kapteyn Astronomical Institute, University of Groningen, PO BOX 800, Groningen NL-9700 AV, the Netherlands
   \and
   INAF Istituto di Astrofisica e Planetologia Spaziali, Via del Fosso del Cavaliere 100, I-00133 Roma, Italy;
   \and 
   Department of Physics and Astronomy, University of Southampton, Highfield, Southampton, SO17 1BJ
   \and 
   Institute of Astronomy, University of Cambridge, Madingley Road, Cambridge, CB3 0HA
   \and
   School of Physics and Astronomy, University of Southampton, Highfield, Southampton, SO17 1BJ, UK
   }
   
   \date{Received date \ Accepted date}

  \abstract
   {Timing analysis of accreting systems is key to probing the structure and dynamics around compact objects. In black-hole low-mass X-ray binaries (BH LMXBs), the compact object accretes matter from a low-mass companion star via Roche-Lobe overflow and forms an accretion disk that occasionally exhibits bright eruptions. The BH LMXB \srclong\ (hereafter \src) underwent one of the brightest outbursts ever recorded in X-rays in August 2023. }
   {We study the timing properties of \src\ in the decaying phase of its outburst based on {\it XMM--Newton} data with a high-time resolution.  
   }
   {We analyzed the power spectrum (PS) and cross spectrum (CS) of \src, which we modeled with Lorentzians. The PS reveals the power distribution of the source across frequencies, and the real and imaginary parts of the CS compare the displacement of the light curves in different energy bands for the different observations. Finally, we simultaneously derived the phase lags and the coherence using a constant phase-lag model. }
   {While the first (soft-state) observation shows no strong variability, the two harder observations exhibit quasi-periodic oscillations (QPOs). Because the QPO is more significantly detected in the imaginary part of the CS than in the PS, we refer to it as the "imaginary QPO". The QPO is more prominent in the soft 0.3--2 keV band than in the hard 2--12 keV band. As the source evolves toward the hard state, the imaginary QPO shifts to lower frequencies, the broadband fractional rms amplitude in the 0.3--2 keV energy band increases, and the rms covariance of the imaginary QPO decreases. 
   Simultaneously, the phase lags increase, and the coherence function drops at the imaginary QPO frequency.}
   {This analysis provides the first type-C QPO detection in a BH XB during the soft--to--hard transition using {\it XMM--Newton} data. The QPO is detected at particularly low energy (0.3--2 keV). Notably, the QPO is significantly detected in the imaginary part of the CS and the PS. Thus, we confirm the physical origin of the coherence drop and the phase-lag excess, which were only observed with {\it NICER} before.}


%

    \maketitle
    \section{Introduction}
    Black hole low-mass X-ray binaries (BH LMXBs) consist of a black hole accreting matter from a low-mass companion star through Roche-Lobe overflow \cite[][]{2002apa..book.....F}. Several of these sources display outbursts, during which their flux increases by a few orders of magnitude over a few weeks to months \citep[][]{2004MNRAS.351..791Z, 2007A&ARv..15....1D, 2020A&A...636A..51H}. The spectral and timing properties of these sources change dramatically throughout the outburst evolution and transition from the Hard state (HS) to the Soft state \cite[SS;][]{2006Remillard_McClintock}. In the HS, the emission from the hot Comptonized medium (often referred to as the "corona") dominates the X-ray spectrum. In contrast, during the SS, a multiblackbody emission from the accretion disk dominates the high-energy spectrum \cite[see][]{1973Shakura_Sunyaev}. The transition between these two spectral states is often referred to as the hard-intermediate state (HIMS) and the soft-intermediate state \citep[SIMS;][]{2005Belloni}. These states are far more rarely observed than the HS and the SS because their lifespan is short \citep{2010MNRAS.403...61D}. \srclong\ (hereafter \src) is one of the few sources that were observed in the decaying phase of the outburst. First detected on August 24, 2023 \citep{2023GCN.34542....1D, 2023ATel16205....1N}, the source emission peaked at around 7 Crab in the 2--20 keV energy band; this event was an exceptionally bright outburst. \src \ is the first BH LMXB that was observed with the Imaging X-ray Polarimetry Explorer \citep[IXPE;][]{2022JATIS...8b6002W} throughout the entire outburst, from the brightening in the hard state \citep[][]{2023ApJ...958L..16V} through the hard--to--soft transition \citep{2024ApJ...968...76I} to the soft state \citep{2024ApJ...966L..35S}, and finally, in the dim hard state \citep{2024A&A...686L..12P}. Optical spectroscopy during quiescence suggests the presence of a black hole primary \citep{2024A&A...682L...1M}, located at $5.5^{1.4}_{1.1}$~kpc \citep{Burridge2025arXiv250206448B}, which defines the system as a low-mass X-ray binary \citep[LMXB;][]{2023ATel16208....1C}. 
    
    \noindent Throughout the outburst, which is visible as a "q" shape in the hardness-intensity diagram \citep[HID; e.g. ][]{Homan_2001, 2004MNRAS.355.1105Fender}, transient BH XBs show fast X-ray variability. 
    Quasi-periodic oscillations \citep[QPOs;][]{1989VanderKlis_QPOs, 1999Psaltis, 2002Belloni, 2004Giannios} are among the components that can be tracked during an outburst. They provide insight into the inner parts of the system. 
    Low-frequency QPOs (LFQPOs) are the most commonly observed events. Their frequencies range from $\approx$0.01 Hz to $\approx$30 Hz \citep{1987Norris, 1998Wijnands, 2004Casella, 2005Belloni}. Based on the strength of the broadband variability and QPO characteristics, LFQPOs can be classified into three main types: type A, B and C \citep{2005ABC}. 
    
    Type-C QPOs are characterized by a strong and narrow peak in the power spectra (PS). The observed fractional root mean square (rms) amplitude ranges from 15\% to 30\%, and the QPO is simultaneously observed with strong broadband variability \citep{2004Casella, 2005Belloni}. The origin of these QPOs remains an open question. Nevertheless, the leading suspects proposed in different models are instabilities in the accretion flow \citep{1999Tagger, 2004Titarchuk, 2009Mondal}, geometrical effects caused by the Lense–Thirring precession \citep{1998Stella, 1999Stella_Luigi, 2009Ingram_Done, 2024MNRAS.528.1668Kubota}, time-dependent Comptonization \citep[][]{2021cosp...43E1695G, 2022A&A...662A.118M, 2022MNRAS.515.2099B}, and a precessing jet \citep[][]{2021NatAs...5...94M}. Type-C QPOs were detected in the HIMS of \src\ with different X-ray telescopes, such as the {\it Neil Gehrels Swift Observatory} \citep{2023ATel16215....1P}, the {\it Neutron Star Interior Composition Explorer} \citep[{\it NICER};][]{2023ATel16219....1D, 2023ATel16247....1B, 2025A&A...697A.229R, 2025ApJ...986....3L}, {\it Insight-Hard X-ray Modulation Telescope} \citep[{\it Insight-HXMT};][]{2024Yu, 2024ApJ...977..148C}, and the {\it Imaging X-ray Polarimetry Explorer} \citep[{\it IXPE;}][]{2024ApJ...961L..42Z, 2024ApJ...968...76I}.\\
    It is crucial to probe the QPO's characteristics during a state transition to understand their origin and formation.
    \cite{Mendez_2024} simultaneously fit the PS and the real and imaginary parts of the cross spectra (CS) using a multi-Lorentzian model in different LMXBs: {\it GX 339--4}, {\it GRS 1915+105}, and during the soft--to--hard transition of {\it MAXI J1820+070}. Their analyses showed that the real and imaginary parts of the CS must be fit because some QPOs can be significant in the CS, but not in the PS. In the soft-energy band (below 1 keV), \cite{2024konig} and \cite{2025A&A...696A.237F} performed a similar analysis of the high-mass X-ray binary (HMXB) {\it Cyg X-1} using {\it NICER} data. Despite the differences in the companion mass, several of these sources (including \src) showed a sudden jump in the phase lags and a coherence drop at a specific frequency. This can be attributed to an additional narrow Lorentzian that is only visible in the imaginary part of the CS. This narrow Lorentzian is likely caused by the variability of the corona. 
    Studies in the rising and decaying phases of the outburst revealed that these features only arise during the decaying phase of the soft--to--hard transition \citep{2022MNRAS.514.2839A, 2023MNRAS.525..854M, 2025bellavita}. This suggests that the inner radius of the disk expands, or that it might come from feedback processes from the corona.
    We present the first timing analysis of \src\ in the soft--to--hard transition with {\it XMM--Newton} data. We used the multi-Lorentzian decomposition of the PS and CS framework presented by \cite{Mendez_2024}.
    From the CS of two different energy bands, we derived the phase lags, which trace the phase angle in the complex Fourier plane of the cross vector for correlated signals \citep{1987ApJ...319L..13Vanderklis, 1989VanderKlis_QPOs, 1997ApJ...483L.115Vaughan, 1999aNowak}. We then calculated the coherence function, which quantifies the linear correlation between the two signals depending on the Fourier frequency \citep{1997ApJ...474L..43Vaughan_Nowak}. \\

    This paper is structured as follows: We present the data set in Section~\ref{data set} and the multi-Lorentzian model fitting of the PS and CS in Section~\ref{Results}. We discuss these results in Section~\ref{Discussion}, together with the phase lags and coherence fitting.

\section{Data Sets}\label{data set}

We observed \src\ with the {\it XMM--Newton European Photon Imaging Camera (EPIC)-PN spectrograph} \citep[covering the 0.15 -- 12 keV energy band;][]{2001XMMJansen, 2001EPIC-PN_Struder} during the soft--to--hard transition \citep[see Table \ref{table:data set}, PID 88500;][]{2020xmm..prop..177C}. The data were processed using the \texttt{Science Analysis Software} \citep[\texttt{SAS};][]{2004ASPC..314..759G} pipeline from ESA. The calibrated photon event files for the {\it PN} camera were produced with \texttt{epproc}. We extracted, filtered for pile-up, and rebinned (0.1~s) the light curves using \texttt{evselect} for all three observations. We defined good time intervals (GTIs) to remove the telemetry gaps (see Fig. \ref{fig: GTIs}), and we computed the power density and cross spectra, which we present in section \ref{Results}.

\section{Analysis and results}\label{Results}
\begin{table*}
\caption{\label{table:PS best fit lor} Best-fitting parameters of the PS and CS for the observations in the soft--to--hard transition. }
    \begin{tabular}{cccccccc}
\hline \hline 
Epoch		&	Comp.	   & $\mathrm{\nu_0}$ [Hz]    & FWHM [Hz]	& $Q$ $\mathrm{= \frac{\nu_0}{FWHM}}$ & $\mathrm{rms_{PS, \ 0.3 - 2 keV}}$ [\%] &	 $\mathrm{rms_{PS, \ 2 - 12 keV}}$ [\%] &  Phase lags [rad]  \tstrut\\ [0.08cm]
\hline																														
			&	L1   & 0.26 $\mathrm{\pm}$ 0.01 & 1.13 $\mathrm{\pm}$ 0.02 &	0.23 $\mathrm{\pm}$ 0.01	& 16.38 $\mathrm{\pm}$ 0.08 &	25.5 $\mathrm{\pm}$ 0.2	& 0.010 $\mathrm{\pm}$ 0.005   \tstrut\\[0.08cm]
XMM2		&	L2   & 2.59 $\mathrm{\pm}$ 0.04 & 1.25 $\mathrm{\pm}$ 0.14 &	2.07 $\mathrm{\pm}$ 0.23	& 3.5 $\mathrm{\pm}$ 0.2 	&	6.3 $\mathrm{\pm}$ 0.7	& 0.70 $\mathrm{\pm}$ 0.05 \\[0.08cm]
			&	L3   & 3.48 $\mathrm{\pm}$ 0.24 & 5.75 $\mathrm{\pm}$ 0.30 &	0.60 $\mathrm{\pm}$ 0.05	& 4.5 $\mathrm{\pm}$ 0.2 	&	15.0 $\mathrm{\pm}$ 0.6	& 	0.26 $\mathrm{\pm}$ 0.03     \\[0.1cm]
				
\hline
$\mathrm{\chi^2/dof}$  &				  	\multicolumn{7}{c}{910.58 / 699 $\approx$ 1.3}\tstrut\\
\hline						   																								
			&	L1   & 0.218 $\mathrm{\pm}$ 0.006 & 0.93 $\mathrm{\pm}$ 0.01 &   0.234 $\mathrm{\pm}$  0.007	& 17.13 $\mathrm{\pm}$ 0.07 &	24.6 $\mathrm{\pm}$ 0.1 & 0.016 $\mathrm{\pm}$ 0.003 \tstrut\\ [0.08cm]
XMM3		&	L2   & 1.98 $\mathrm{\pm}$ 0.01 	& 0.48 $\mathrm{\pm}$ 0.05 &   4.12  $\mathrm{\pm}$  0.43 	& 2.9 $\mathrm{\pm}$ 0.1 	&	3.7 $\mathrm{\pm}$ 0.3  & 0.81  $\mathrm{\pm}$ 0.04  \\[0.08cm]
			&	L3   & 2.48 $\mathrm{\pm}$ 0.08 	&	4.6 $\mathrm{\pm}$ 0.1 &   0.54  $\mathrm{\pm}$  0.02  	& 6.4 $\mathrm{\pm}$ 1.4   	&  17.3 $\mathrm{\pm}$ 0.2  & 0.32  $\mathrm{\pm}$ 0.01  \\
\hline 
$\mathrm{\chi^2/dof}$ &		 				\multicolumn{7}{c}{1338.33 / 699 $\approx$ 1.9}\tstrut\\
\hline
    \end{tabular}
        \tablefoot{
``Comp'' refers to the model components and L1-3 refer to the three fit Lorentzians (defined in equations \ref{eq: lorentz_ps} for the PS and in equation \ref{eq: lorentz_cs} for the CS).
}
\end{table*}

\begin{figure*}
    \centering
   \includegraphics[width=0.45\textwidth]{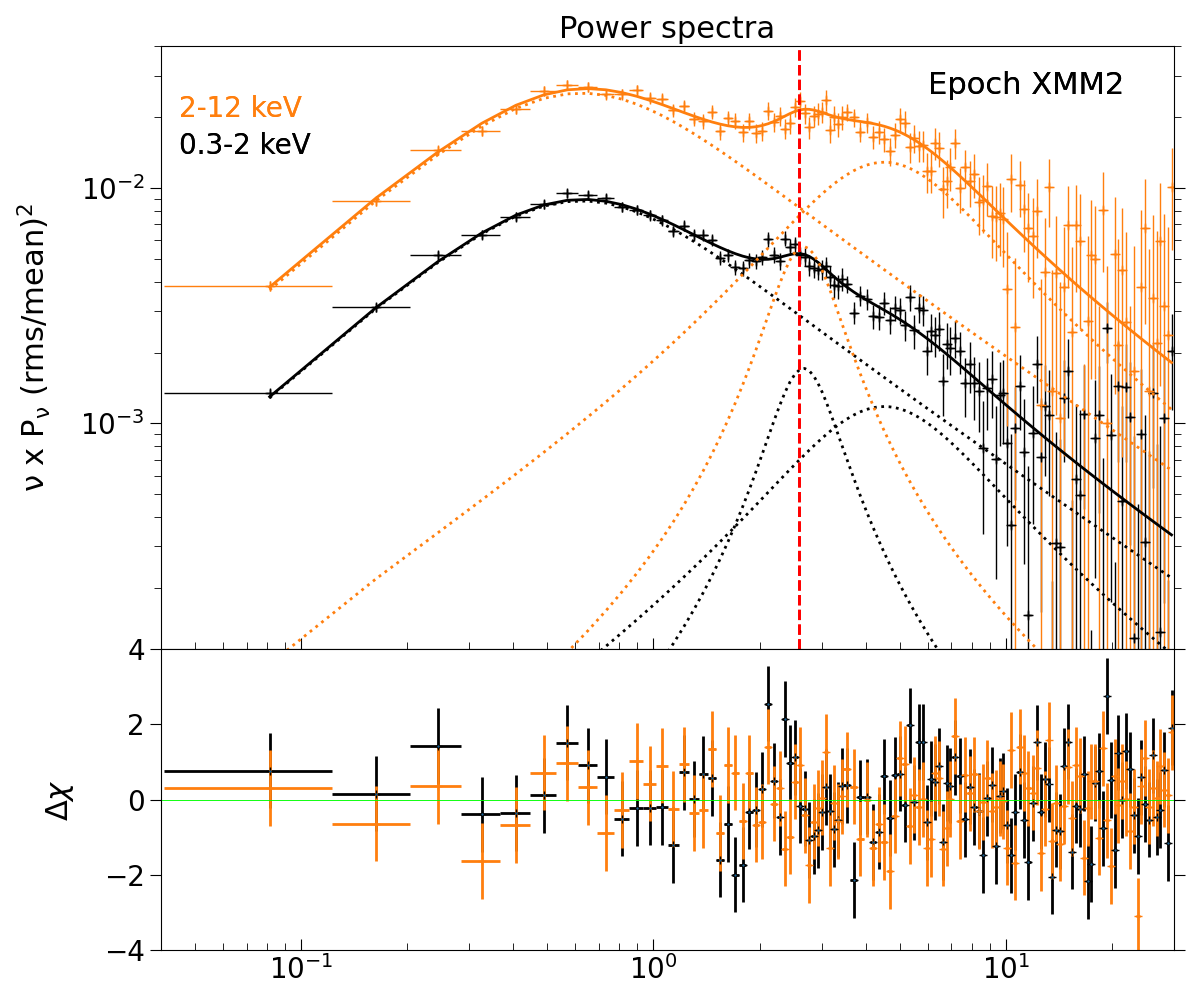}  
    \includegraphics[width=0.45\textwidth]{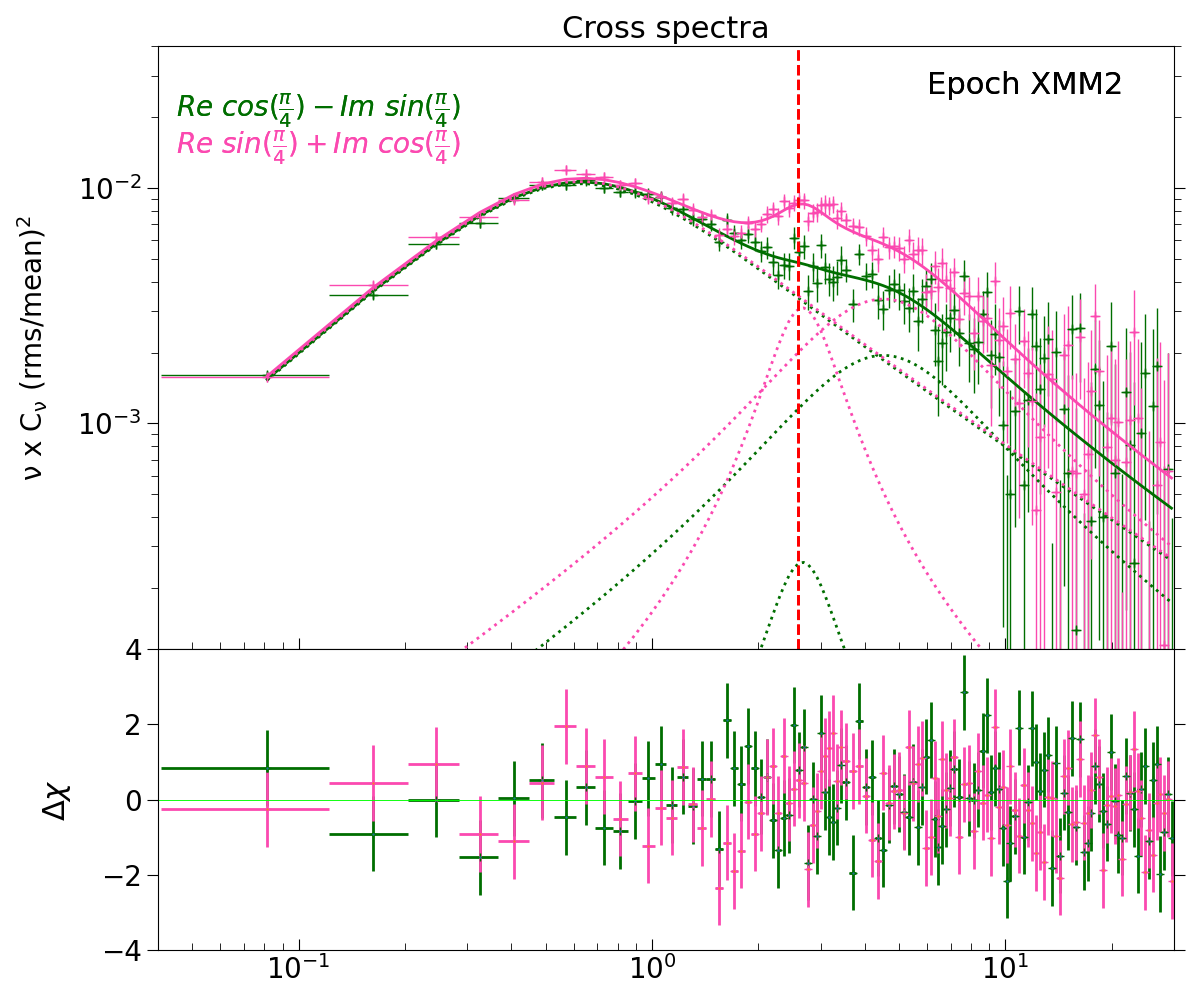}
    
    \includegraphics[width=0.45\textwidth]{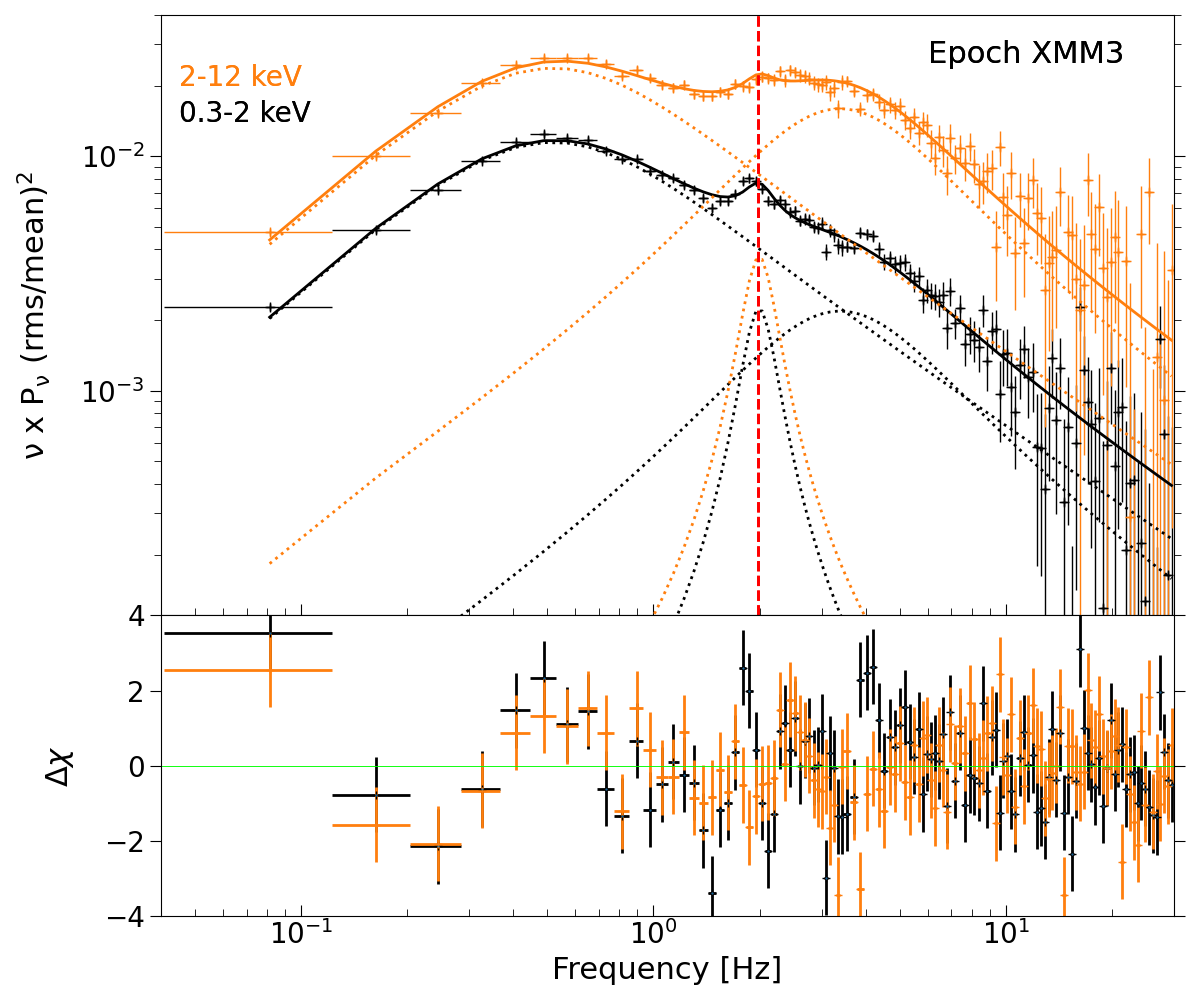}
    \includegraphics[width=0.45\textwidth]{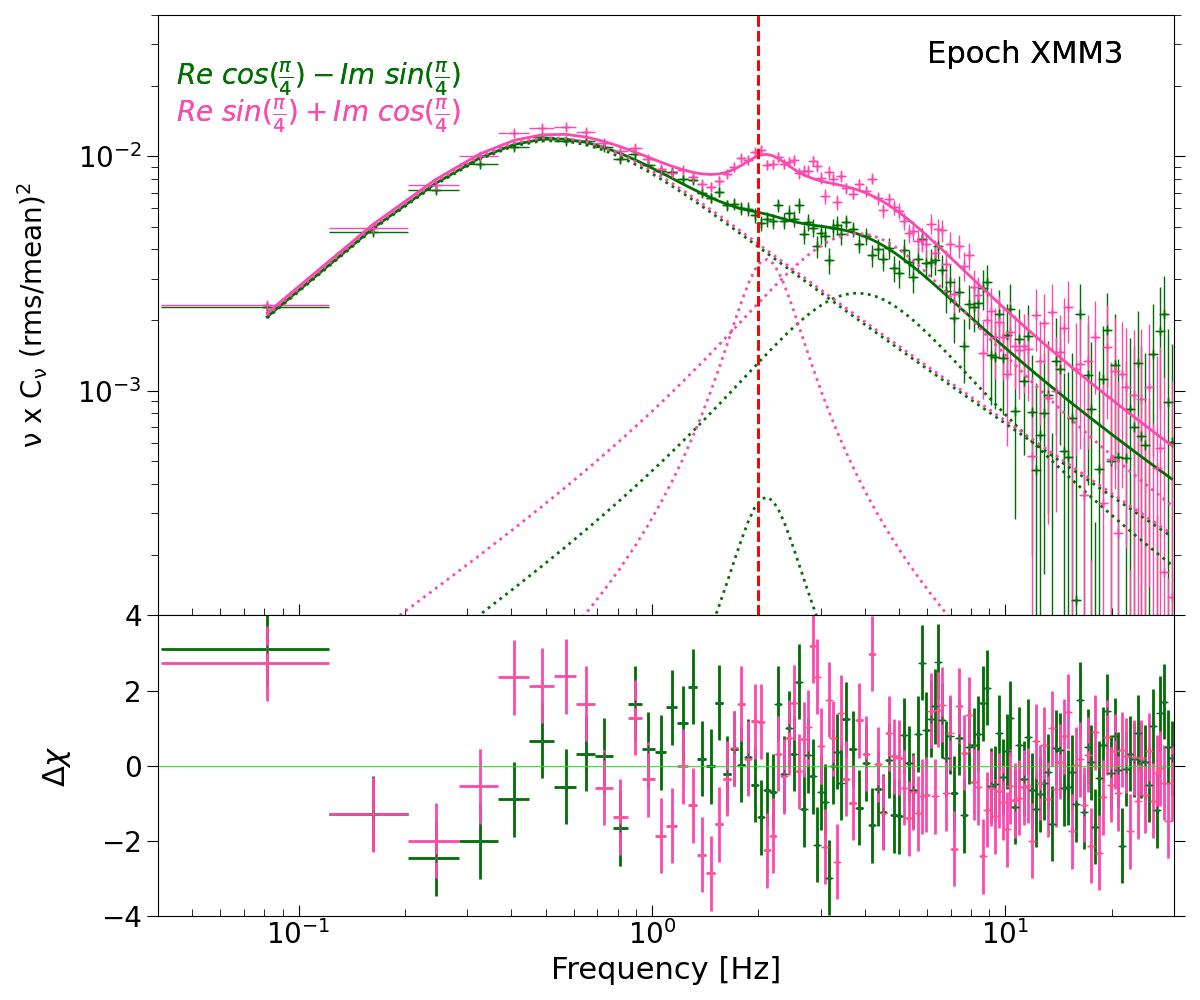}
    \caption{\textit{Left panel}: Power spectra of the second (top) and third (bottom) EPIC-PN observations of \srclong\ in the 0.3--2 keV (black) and 2--12 keV (orange) energy bands. \textit{Right panel}: real (green) and imaginary (pink) parts of the cross spectra of the same two observations using the 0.3--2 keV and 2--12 keV energy bands, rotated by 45\degree (see Sect. \ref{real and imaginary parts of the CS} for more details). The vertical red line shows the position of the central frequency $\nu_0$ of the imaginary QPO. The total model is shown with the continuous line, and the individual Lorentzians are shown as dotted lines. The residuals defined as $\mathrm{\Delta \chi=}$ (data--model)/error are plotted below each fit. }
    \label{fig: PS fit}
\end{figure*}

Using the \texttt{General High-Energy Aperiodic Timing Software (GHATS\footnote{\url{http://astrosat-ssc.iucaa.in/uploads/ghats_home.html}})}, we calculated the PS in the 0.3--2 keV and 2--12 keV energy bands and the cross spectra between those same two bands. We fixed the sampling time to 0.006 s ($200\times$ the native instrumental resolution), which defines the Nyquist frequency at $\mathrm{0.5/0.006 = 83.3}$ Hz. Based on the typical length of the GTI segments allowed by the telemetry gaps, we set the interval length for computing the PS to $\mathrm{2048 \ \times \ 0.006 s  \approx 12.3 \ s}$, which corresponds to a minimum frequency of $\approx$0.08~Hz. Consequently, the second observation (ObsID 0885000701) had 641 segments, and the third observation (ObsID 0885000801) had 1605 segments. To increase the signal-to-noise ratio at high frequencies, we logarithmically rebinned the PS and CS by a factor of $\mathrm{10^{0.01} \approx 1.023}$. The PS and CS were normalized to fractional rms units \citep[][]{1989VanderKlis_QPOs}. The Poisson noise and cross-talk levels were estimated empirically from the 30--80 Hz frequency range, where no signal was detected, and they were subtracted from the PS and the real part of the CS. Therefore, the PS and CS were only fit up to 30 Hz. Using the PS and CS, we computed the phase lags and the coherence for all observations, and we then fit them with the \texttt{X-Ray Spectral Fitting Package \citep[Xspec;][]{1996ASPC..101...17A}}.

\subsection{The power spectra}

No significant variability was observed in the PS of the first (soft state) observation, as expected. This analysis, therefore, focuses on the last two observations during the soft--to--hard transition.
The PS were computed using \texttt{GHATS} and were fit in \texttt{Xspec} with an additive multi-Lorentzian model for each observation (see Fig. \ref{fig: PS fit}). Following \cite{Mendez_2024}, the Lorentzians were assumed to be coherent between the two energy bands across the Fourier frequencies, but incoherent between each other. Thus, the PS in two energy bands, x and y, can be defined as a sum of Lorentzians,

\begin{equation}\label{eq: lorentz_ps}
\begin{split}
    G_{xx} (\nu) = \sum_{i=1}^3 G_{xx;\ i} (\nu) = \sum_{i=1}^3 A_i \ L(\nu; \nu_{0;i}, \Delta_i)\\
    G_{yy} (\nu) = \sum_{i=1}^3 G_{yy; \ i} (\nu) = \sum_{i=1}^3 B_i \ L(\nu; \nu_{0;i}, \Delta_i)\ ,
\end{split}
\end{equation}
with $\mathrm{\textit{i}=1,\ 2,\ 3}$ for the three Lorentzians, $A_i$, $B_i$ are the integrated power from zero to infinity of the Lorentzians in the band x and y, respectively, $\mathrm{\nu_{0;\textit{i}}}$ is the central frequency for each individual Lorentzian, and $\mathrm{\Delta_{\textit{i}}}$ is their full width at half maximum (FWHM). Since for each Lorentzian function we assumed the coherence function to be $\mathrm{\gamma^2_{\textit{xy;i}} (\nu) = 1}$ and the modulus square of the CS to be $\mathrm{|G_{\textit{xy}} (\nu)|^2 =}$ $A_i$ $B_i$ $\mathrm{L^2(\nu; \nu_{0;\textit{i}}, \Delta_{\textit{i}})}$, the former is defined as
\begin{equation}
    \gamma^2_{xy} (\nu) = \frac{|G_{xy} (\nu)|^2}{G_{xx} (\nu)G_{yy} (\nu)}  \ .
\end{equation}
Table \ref{table:PS best fit lor} gives the best-fitting parameters from the \texttt{Xspec} fitting with three Lorentzians for each observation. The central frequency $\mathrm{\nu_{0}}$ and the FWHM of the Lorentzians in the 2--12 keV energy band are linked to the values from the low-energy band (0.3--2 keV). Only the fractional rms amplitude remains a free parameter in both bands. The quality factor $Q$ is defined as the ratio of the central frequency of the QPO and the FWHM.
The best-fits using the multi-Lorentzian model give $\mathrm{\chi^2 \ / \ dof \approx 1.3}$ and $\mathrm{\chi^2 \ / \ dof \approx 1.9}$ for the second and third observation, respectively. The model describes the data correctly in both cases, without prominent features in the residuals. 
The PS has a greater rms amplitude at higher energy overall, with a total rms of $\approx$17\% in 0.3--2 keV versus $\approx$30\% in 2--12 keV. \\
\indent Additionally, the rms of the first and last Lorentzians (referred to as L1 and L3) increase in the 0.3--2 keV energy band as the outburst evolved from the second to the third observation: the rms increases from $\approx$16.4\% to $\approx$17.1\% for L1 and from $\approx$4.5\% to $\approx$6.4\% for L3. In the 2--12 keV energy band, the rms similarly increases for L3, but decreases for L1 from $\approx$6.3\% to $\approx$3.7\%. On the other hand, the rms amplitude of L2 decreases in both energy bands from $\approx$3.5\% to $\approx$2.9\% in the 0.3--2 keV energy band and from $\approx$6.3\% to $\approx$3.7\% in the 2--12 keV energy band.   \\
\indent Furthermore, the central frequencies of the three Lorentzians shift toward lower frequencies as the source hardens. They decrease by $\approx$16\%, $\approx$23\%, and $\approx$28\% for L1, L2, and L3, respectively, between the second and third observation. The Lorentzians also become narrower, with a decrease in the FWHMs of $\approx$18\%, $\approx$62\% and $\approx$20\% between the second and third observation (see Table \ref{table:PS best fit lor}). As a result, when the source hardens, the quality factor of the second Lorentzian increases from 2.07 to 4.12, but decreases for L3 from 0.60 to 0.54. The $Q$ factor of L1 simultaneously remains constant around 0.23 within the errors.

\subsection{The real and imaginary parts of the cross spectrum}\label{real and imaginary parts of the CS}

Following \cite{Mendez_2024}, the real and imaginary parts of the CS were fit in a similar way as the PS. They were decomposed into a linear combination of Lorentzians multiplied by the phase-lag dependence using the relation
\begin{equation} \label{eq: lorentz_cs}
\begin{split}
    \Re[G_{xy} (\nu)] = \sum_{i=1}^3 C_i \ L(\nu; \nu_{0;i}, \Delta_i) \ \cos{(\Delta\phi_{xy;i} (\nu) + \pi/4)}\\
    \Im[G_{xy} (\nu)] = \sum_{i=1}^3 C_i \ L(\nu; \nu_{0;i}, \Delta_i) \ \sin{(\Delta\phi_{xy;i} (\nu) + \pi/4)} \ ,
\end{split}
\end{equation}
with $C = \sqrt{AB}$, and the frequency-dependent phase lags between the two signals were defined as
\begin{equation}
   \Delta\phi_{xy;i} (\nu) = \tan^{-1} \left(  {\frac{\Im[G_{xy;i} (\nu)]}{\Re[G_{xy;i} (\nu)]}} \right) = g_i(\nu;p_{j,i}) \ .
\end{equation}
We explicitly treated the phase lags of  the \textit{ith} Lorentzian as a function of the parameters $p_{\textit{j,i}}$. 
From now on, we assume the phase lags of each Lorentzian to be constant with frequency, that is, $g_i(\nu;p_{j,i}) = \Delta\phi_{\textit{i}}$, where $\Delta\phi_{\textit{i}}$ is a parameter of the model. This is the constant phase-lag model as defined by \cite{Mendez_2024}. The phase lags of \src\ are close to zero across the Fourier frequencies, and the amplitude of the imaginary part of the CS is therefore smaller than that of the real part. Consequently, the cross vectors can be rotated by 45\degree \ to have comparable real and imaginary parts. This generates more stable fits in \texttt{Xspec} and does not affect the best-fitting parameters \cite[see][]{Mendez_2024}. By multiplying the cross vector (decomposed into a real and imaginary part) by the rotation matrix, we therefore calculated the real and imaginary parts of the CS as $\mathrm{\Re[\cos{(\pi/4)}] \ - \ \Im [\sin{(\pi/4)}]}$ and $\mathrm{\Re[\sin{(\pi/4)}] \ + \ \Im[\cos{(\pi/4)}]}$, respectively. 
The central frequencies and FWHMs of the three Lorentzians in the CS are tied to those in the 0.3--2 keV band of the PS. The normalization of the real part of the CS is given by the parameter $C$, as previously defined. 
Table \ref{table:PS best fit lor} shows the best-fitting parameters of the phase lags of the CS between 0.3--2~keV and 2--12~keV. The rotated real (green) and rotated imaginary (pink) parts of the CS we fit are shown in the right panel of Fig.~\ref{fig: PS fit}. 
Below 1.5 Hz, the rotated real and imaginary parts of the CS overlap in the two energy bands for both observations. 
Above 1.5 Hz, the rotated imaginary part of the CS in the 2--12~keV energy range is larger than the rotated real part. A bump is observed in the rotated imaginary part of the CS at $\approx$3~Hz, and it is well fit by a narrow Lorentzian. In both observations, we identify a narrow Lorentzian in the PS and the CS with a larger imaginary component than in the real part. We call this QPO the imaginary QPO, as first expressed by \cite{Mendez_2024}. While the rms of the imaginary QPO stays more or less constant when the system hardens, the real part of the QPO increases.

\begin{figure*}
    \centering
   \includegraphics[width=0.39\textwidth]{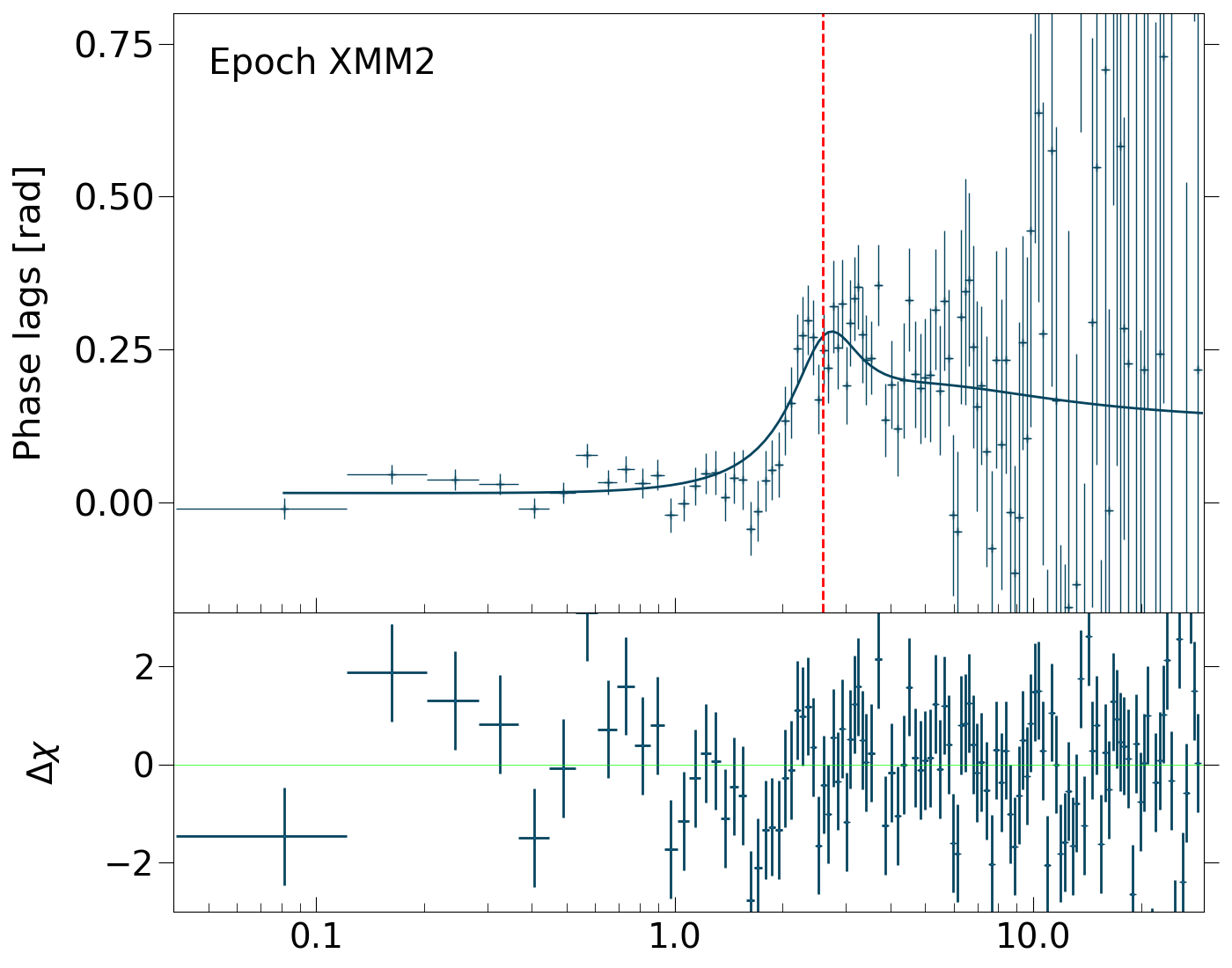} 
   \includegraphics[width=0.39\textwidth]{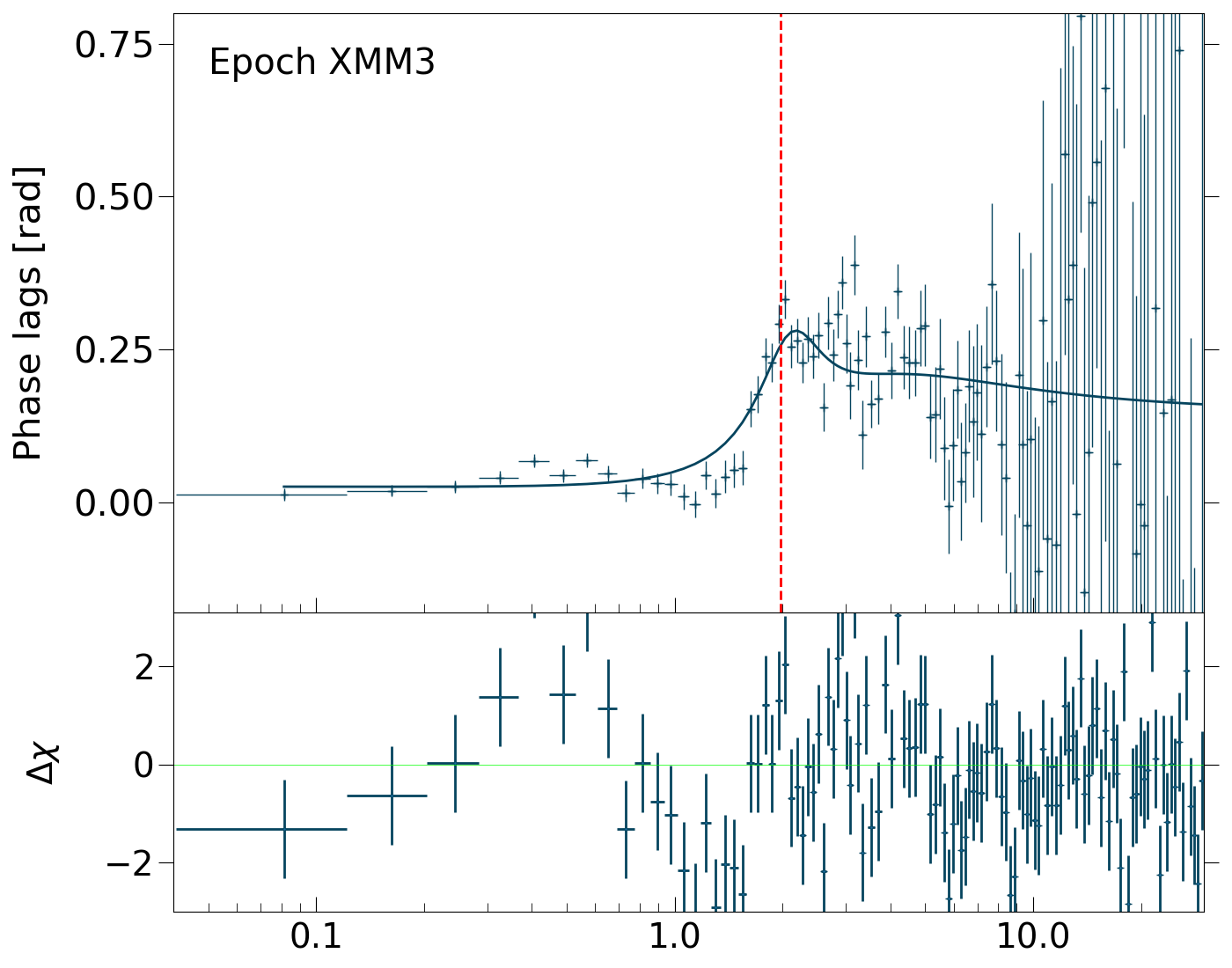}
   
   \includegraphics[width=0.39\textwidth]{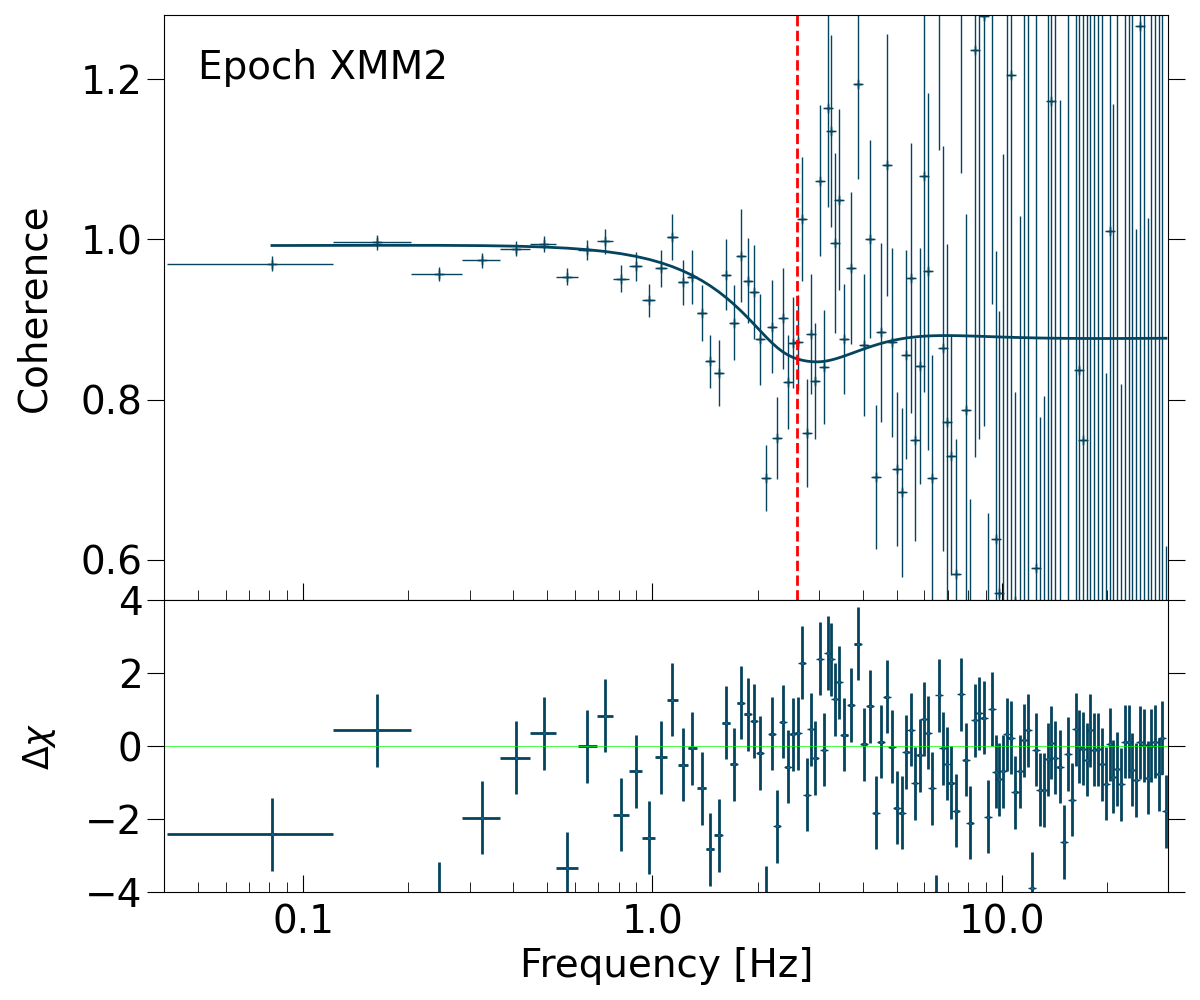}  
    \includegraphics[width=0.39\textwidth]{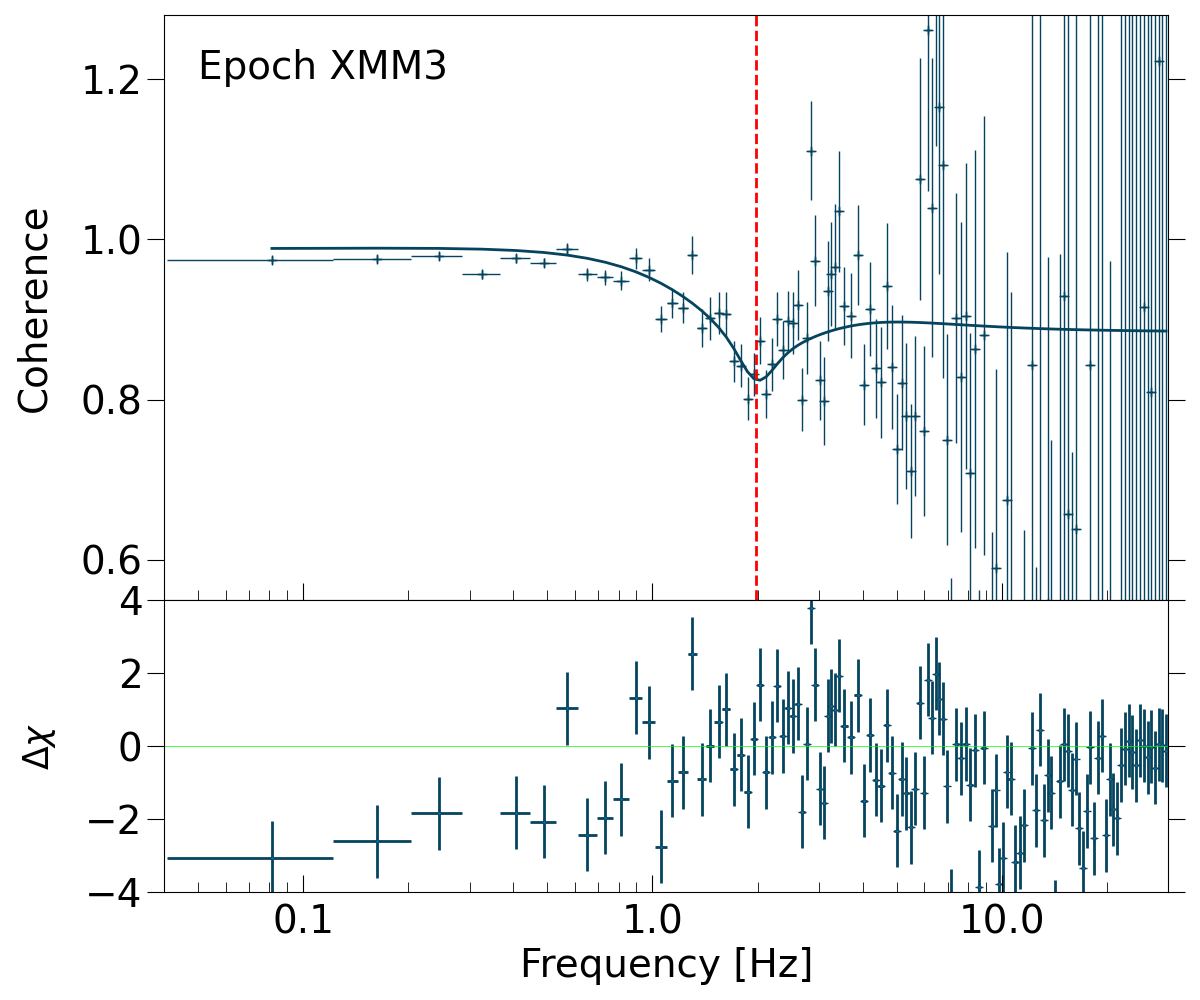}
    \caption{Phase lags in radian (top) and the coherence function (bottom) for the second (left) and third (right) EPIC-PN observation of \srclong. The continuous lines show the predicted model from the fits of the PS and CS represented in Fig. \ref{fig: PS fit}.
    The bottom panels show the residuals defined as (data-model)/error. The dotted vertical red line indicates the central frequency of the imaginary QPO for each observation. }
    \label{fig: Phase lags and coherence fit}
\end{figure*}

\subsection{The phase lags and coherence}
The last column of Table~\ref{table:PS best fit lor} shows the best-fitting parameters for the phase lags. The phase lags of the three Lorentzians increase by $\approx$37\%, $\approx$13\%, and $\approx$19\% (for L1, L2, and L3, respectively) between the second and third observation. From the PS and CS fits, we derived a model of the phase lags and the coherence, calculated in the same energy bands as the power and cross spectra.
Fig.~\ref{fig: Phase lags and coherence fit} displays the phase lags (top panel) and the coherence (bottom panel) for the two observations, together with the predicted model. Below 1~Hz, the phase lags are constant around $\approx$0~rad on average, with a slight excess at $\approx$0.2~Hz and $\approx$0.5~Hz that is significant at the 68\% confidence level. Then, the phase lags reach a maximum value at the imaginary QPO frequency, marked by the vertical dotted red line, and stabilize to a higher constant value at $\approx$1.8~rad after this for both observations. In the second observation, however, the phase lags are narrower and are dominated by the uncertainties above 7~Hz for both observations. \\
\indent While the phase lags increase at the central frequency of the imaginary QPO, the coherence drops. Below 1~Hz, the average coherence value is $\approx$1, it then drops to $\approx$0.8 at the frequency of the imaginary QPO and remains constant at $\approx$0.9 after the dip. Similarly to the phase lags, the residuals drop significantly at $\approx$0.5~Hz, aligning with a slight decrease in the coherence. These features suggest that an additional component contributes to the variability at low frequency. We did not add another Lorentzian to account for this component in our model, because we aimed to fit the prominent variable components with as few Lorentzians as possible.

\subsection{The energy dependence of the QPO lags and rms amplitude}
\begin{figure}
    \centering
    \includegraphics[width=0.45\textwidth]{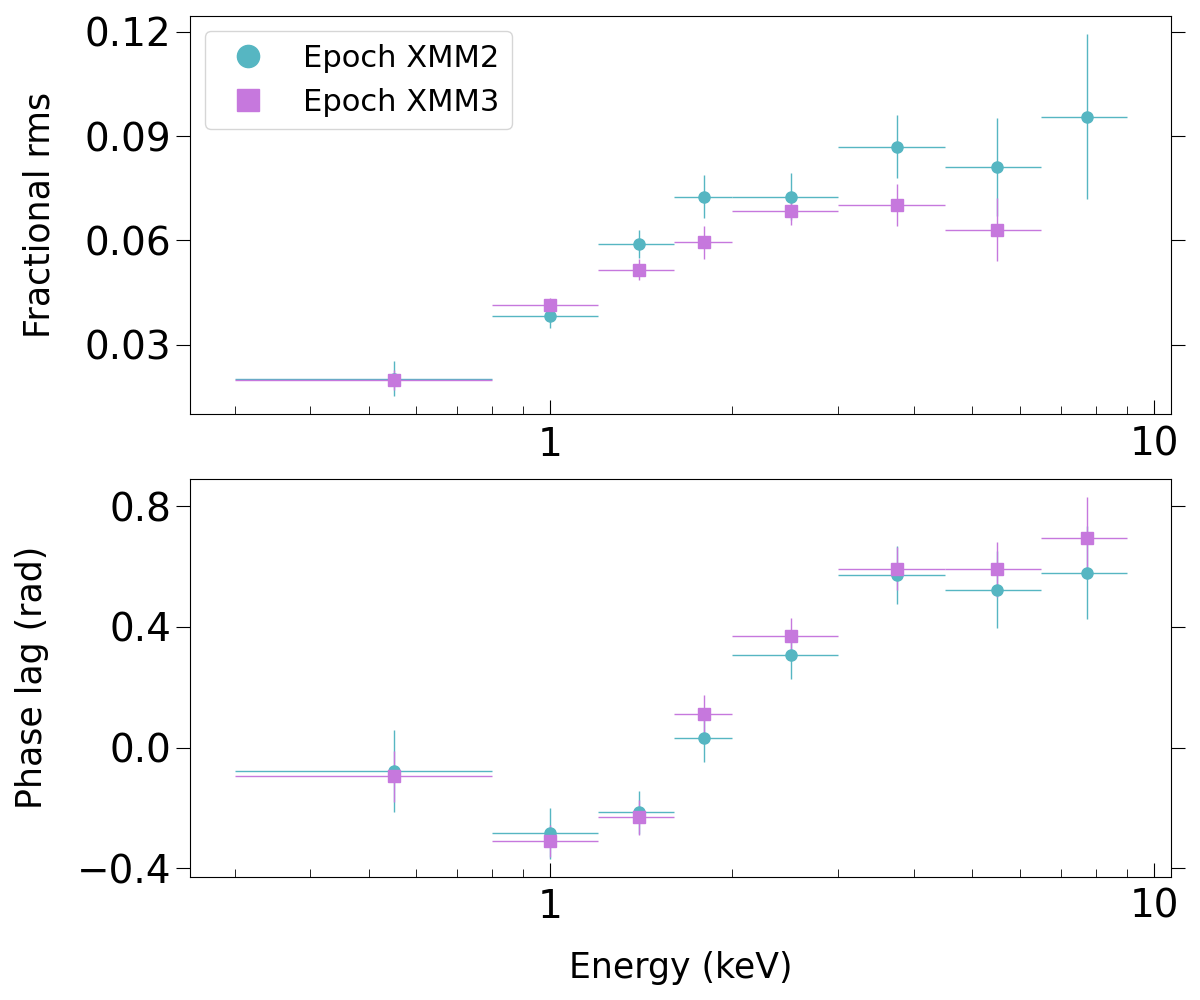}   
    \caption{ Energy dependence of the fractional rms (top) and the phase lags (bottom) of the imaginary QPO for the second (blue circles) and third (purple squares) EPIC-PN observations of \srclong. }
    \label{fig: Energy rms and lags}
\end{figure}

To analyze the QPO lags and rms amplitude energy dependence, we computed the fast Fourier transforms (FFTs) and CS for each narrow energy band relative to the full (0.3--12~keV) energy band at the imaginary QPO frequency. The data were then fit using the multi-Lorentzian model we defined above, and the central frequencies and FWHMs were fixed to the best-fit values listed in Table \ref{table:PS best fit lor}. Fig. \ref{fig: Energy rms and lags} shows the energy dependence of the fractional rms and the phase lags.
In epoch XMM2, the rms amplitude of the QPO increased from 2\% to 9.5\% between 0.5~keV and 8~keV. In epoch XMM3, the rms amplitude increased from 2\% to $\approx$7\% between 0.5~keV and 5.5~keV. The higher-energy bands are poorly constrained and the errors are significant up to 3$\sigma$. We therefore did not consider them in the analysis. \\
On the other hand, the phase lags reached a minimum value at 1~keV. In epoch XMM2, the phase lags decreased from $\approx$ --0.08 rad to $\approx$ --0.28 rad between 0.5 keV and 1 keV and increased again from $\approx$ --0.21 rad at 1.2 keV to $\approx$0.58 rad at 8 keV. In epoch XMM3, a similar behavior is observed with a decrease from $\approx$ --0.09 rad at 0.5 keV to a minimum value of $\approx$ --0.31 rad at 1 keV that increased above 1 keV.

\section{Discussion and conclusions}\label{Discussion}
In section \ref{Results}, we showed that the  {\it XMM EPIC-PN} data of the BH LMXB \srclong\ reveal a type-C QPO in the soft--to--hard transition. While the characteristic Lorentzians frequencies shift to lower values as the source hardened, the phase lags increased and the coherence function dropped at the QPO frequency. This supports again the possible cause--to--effect relation between the type-C QPO and the features observed in the lags and the coherence. An in-depth discussion is provided below.

\subsection{The type-C QPO}
We identified a narrow Lorentzian in the PS characterized by a higher rms amplitude in the imaginary part than in the real part of the CS, where the rms of the real counterpart increased between the second and the third observation. The origin of the imaginary QPO is still unclear: In {\it Cyg X-1}, \cite{2024konig} suggested a beat between the two broad Lorentzians, that might create a third component causing the QPO, as is often observed in the PS, whereas \cite{Mendez_2024} argued that it comes from an independent phenomenon. Nonetheless, the imaginary QPO of the source seems to correspond to the QPO observed in the PS.\\
\indent The QPO identified in \src's PS and CS has a central frequency at $\approx$2.6~Hz in the second observation that shifts to $\approx$2~Hz in the third observation. Furthermore, the total rms is $\approx$17\% and $\approx$18\% in the 0.3--2 keV energy band for the second and last observation, respectively, and it reaches $\approx$30\% in the 2--12 keV energy band for both observations.
Based on the brightness of \src\ and on high-rms LFQPO in the PS and CS, we classified the LFQPO that is observed in the soft--to--hard transition as a type-C QPO \citep[][]{2011MNRAS.418.2292M}.
In the rms–intensity diagram for black hole transients defined by \cite{2011MNRAS.410..679M}, the source would be close to the hard line, which is defined as a linear relation between the absolute rms and the count rate observed in the canonical low-hard state. Weak type-C QPOs are generally observed at the hard line in the hard state and migrate above the line as the rms increases and reaches 30\%, and they progressively leave the hard line. The $Q$ factor values for \src\ are lower than those usually observed for type-C QPOs ($\approx$7--10), however. Low $Q$ factors in type-C QPOs were also observed in {\it MAXI J1348--630} \citep{2022MNRAS.514.2839A} and {\it MAXI J1820+070} \citep{2023MNRAS.525..854M}, where they ranged between 0.2 and 7. \\
\indent On the other hand, as the system hardened, the QPO moved to lower frequencies. This frequency shift (probably induced by the Comptonizing plasma that dominates the emission in the hard state) has previously been observed in other XBs \cite[{\it MAXI J1820+070};][and {\it Cyg X-1}; \citealp{2024konig, 2025A&A...696A.237F}]{Mendez_2024, 2025bellavita}. 
Furthermore, during the soft--to--hard transition of {\it MAXI J1348--630}, \cite{2025Alabarta} interpreted the shift in the QPO frequency to lower frequencies as indicative of coronal expansion. 
In the case of {\it GRS 1915+105}, \cite{2022NatAs...6..577M} and \cite{2022MNRAS.513.4196G} explained the frequency shift by a perpendicular contraction of the corona with respect to the accretion disk. Thus, the frequency shift suggests a geometric change in the corona.\\
Additionally, observed in the soft--to--hard transition, the type-C QPO can be attributed to the expansion of the truncated inner radius of the disk or to feedback processes from the corona \citep{2022MNRAS.515.2099B}. The Lense-Thirring precession, in which the inner accretion flow precesses with respect to the thermal disk \citep[see][]{2009Ingram_Done}, has also been proposed as the origin of type-C QPOs in XBs. This was supported by studies that reported a tight anticorrelation between the outer radius of the Comptonizing region and $\mathrm{\nu_{QPO}}$ \citep[][]{2024MNRAS.528.1668Kubota}. The analysis of {\it GRS 1915+105} by \cite{2022MNRAS.511..255N} revealed an inner disk radius that was too small and a reprocessing time that was too long to support this model, however. \\

\indent In addition to the type-C QPO, the broadband Lorentzians L1 and L3 similarly shift to lower frequencies as the source hardens. In the 0.3--2~keV energy band, however, their rms amplitude increases as the source transitions, unlike that of the type-C QPO. This behavior was previously observed in {\it GX 339-4} \citep[][]{2011MNRAS.410..679M}. In the 2--12~keV hard band, the rms increases for L3 for a harder emission, but it decreases for L1. Thus, as suggested by \cite{2025bellavita}, the two broad Lorentzians seem to be independent from each other and might come from variability processes that differ from the QPO. 
\cite{2024konig} supported this assumption by proposing that L1 comes from the accretion disk and is modulated by the Comptonization and that L3 is associated with the Comptonized medium. On the other hand, \cite{2009MNRAS.397..666W} linked the L1 variability to an unstable accretion flow in the disk (although they reached this conclusion by fitting the covariance spectrum with the same model as the time-averaged spectrum, which only holds when the overall normalization of the model alone varies).

\subsection{Phase-lag jump}
At the QPO frequency, the phase lags reach a maximum, also called the cliff. This was previously observed in other XBs \citep{Mendez_2024, 2025bellavita}. Before and after the QPO frequency, however, the phase lags remain constant at $\approx$0~rad and $\approx$0.18~rad, respectively, for both observations, with a sharper cliff in the third observation. In the frequency range in which one Lorentzian dominates the PS, the phase lags remained constant on average, where each Lorentzian had its own lags, potentially caused by separate resonances \citep{Mendez_2024}. At the intersection of the two Lorentzians, in contrats, the phase lags transition from one constant value to the next.
\cite{1999bNowak} attributed this difference to the total phase lags at each Fourier frequency being equal to the average phase lags of the individual Lorentzians, weighted by the product of their amplitudes in the two energy bands at which they were calculated \cite[defined as the small-angle approximation of equation 9 in][]{Mendez_2024}. As a result, the phase lags are constant over a given frequency range when the Fourier amplitudes of one Lorentzian dominate. \\
Moreover, no cliff is observed in the hard-energy bands. Similarly observed in the rising phase of the outburst of \src, \cite{2024Yu} did not find a phase-lag jump in the 1--10~keV and 25--150~keV energy bands, but still detected it in the soft energy bands, 1--2~keV and 2--10~keV. This might suggest that soft energies are essential to observe the phase-lag jump, regardless of the transition direction. This similarly indicates
that the QPO originated from variations in a soft component that is linked to the accretion disk. Furthermore, as the source hardens, the Comptonization-driven variability from the corona intensifies and might introduce additional variability in the accretion disk.

\subsection{Coherence drop }
While the phase lags reached a maximum value at the QPO frequency, the coherence reached a local minimum \citep{2024konig, Mendez_2024, 2025bellavita, 2025A&A...696A.237F}, with different plateau values before and after the QPO frequency. Before the drop, the constant coherence of $\approx$1 indicates the physical interconnection between the variability in the disk and the corona \citep{2024konig}. After the drop, the constant value of $\approx$0.9 indicates an increasing variability of the corona at the QPO frequency. 
In order to test this assumption, we fit the PS, CS, phase lags, and coherence with only two Lorentzians instead of three. We obtained worse fits, with a $\mathrm{\chi^2 / dof \ \approx 1.5}$ for the second observation, and a $\mathrm{\chi^2 / dof \ \approx2.4}$ for the third observation. In the residuals of the PS and CS, we identified a broad Gaussian-shaped component that was significant between 3$\sigma$ and 4$\sigma$ for the phase lags and coherence fits. This indicates that the model misses a component. Consequently, three Lorentzians (i.e., including a narrow one), are required at least for the model to correctly fit the PS, CS, phase-lag cliff, and coherence drop of \src, similarly to {\it MAXI J1820+070} \citep{2025bellavita, 2025A&A...696A.237F} and {\it Cyg X-1} \citep{2024konig}. Ultimately, the drop in coherence might come from a weaker interaction between the large corona and the accretion disk at the QPO frequency as the transition evolves \citep{2025Alabarta}, which would explain the sharper coherence drop and phase-lag cliff observed in the last observation (see Fig. \ref{fig: Phase lags and coherence fit}).

The phase-lag jump and the coherence drop seem to be generated overall by low-energy variations from a soft component in the accretion disk at the type-C QPO frequency, induced by variations in the corona geometry \citep{2022NatAs...6..577M, 2025Alabarta}. Only observed in the decaying phase of the outburst \citep{Mendez_2024, 2025Alabarta, 2024konig, 2025bellavita, 2025A&A...696A.237F}, these features are significant at low X-ray flux in the soft--to--hard transition. Moreover, the phase lags and coherence of \src\ show a significant excess at $\approx$0.4~Hz and $\approx$3~Hz that was previously observed in other sources with a different instrument \citep{Mendez_2024, 2024konig, 2025ApJ...990...43R}. This confirms the physical origin of the signal.

\subsection{QPO energy dependence}
In \src, the QPO frequency does not depend on energy. It stays constant in the 0.3--2~keV, compared to the 2--12~keV energy bands. At the QPO frequency ($\approx$2 Hz), the fractional rms continually increases with energy as a result of the decreasing contribution of the disk at higher energies \citep[as observed in other sources;][]{2022MNRAS.514.2839A, 2025Alabarta, 2025A&A...696A.237F, 2025bellavita}. \cite{2022MNRAS.515.2099B} showed that the fractional rms as a function of energy is strongly dependent on the power-law index ($\mathrm{\Gamma}$), based on the variable Comptonization model for low-frequency QPOs, \textit{vKompth}. Similarly, the rms of \src\ decreases as the source hardens (i.e., $\mathrm{\Gamma}$ decreases), this suggests that the drop in the QPO rms amplitude may be driven by the decrease in $\mathrm{\Gamma}$. 
The phase lags simultaneously reach a minimum value around 1~keV for \src, around 1.6~keV for {\it MAXI J1348--630} \citep[][]{2025Alabarta}, and around 1.5~keV for {\it MAXI J1820+070} \citep[][]{2023MNRAS.525..854M}, and they remain constant $\gtrapprox$4~keV. The U-shaped energy lags might be explained by a fraction of the cold seed photons from the accretion disk that enters the extended corona in the horizontal direction and results in a significant drop in the phase lags \citep{2025ApJ...986....3L, 2025Alabarta, 2025bellavita}. Furthermore, as the source hardens, \cite{2025bellavita} reported a stronger drop in the phase lags. 

\section{Summary}
\begin{itemize}
  \renewcommand\labelitemi{-}
    \item This analysis revealed a type-C QPO in the soft--to--hard transition of the LMXB \srclong. This represents the first ever detection of a type-C QPO in the soft--to--hard transition of a BH XB using {\it XMM--Newton} data (it was previously only observed with other telescopes, e.g., {\it NICER} or {\it RXTE}).\\
    
    \item The QPO is significantly detected in the imaginary part of the CS and the PS, with a central frequency at $\approx$2.6 Hz, shifting to $\approx$2 Hz as the source spectrum hardens. This indicates geometrical changes in the corona. The broadband fractional rms amplitude in the 0.3--2 keV energy band increases simultaneously and the rms covariance of the QPO decreases, which suggests fluctuations in the accretion disk (probably from the mass-accretion rate) and modulated by the Comptonization. Considering the total rms of about 17\% and 30\% between the last two observations and the brightness of the source, we classified the QPO as a type-C QPO.\\
    
    \item The unique detection of the coherence drop and the phase-lag excess at the QPO frequency in \src\ suggests a weaker interaction between the large corona and the accretion disk at the QPO frequency as the transition evolves from the soft to the hard state.
    The first detection of the coherence drop in the soft--to--hard transition with a telescope other than {\it NICER} implies that it has a physical origin and is not an instrumental artefact. 
\end{itemize}

\noindent Follow-up modeling of this source may help us to understand the relative strength of the type-C QPO in the imaginary component of the CS.

\begin{acknowledgements}
        Research is based on observations obtained with {\it XMM-Newton}, an ESA science mission with instruments and contributions directly funded by ESA Member States and NASA. 
        M.B. acknowledges the support from GAUK project No. 102323. J.S. thanks GACR project 21-06825X for the support. NCS acknowledges support from the Science and Technology Facilities Council (STFC) grant ST/X001121/1. FG acknowledges support from PIBAA 1275 and PIP 0113 (CONICET). FG was also supported by grant PID2022-136828NBC42 funded by the Spanish MCIN/AEI/10.13039/501100011033 and ERDF A way of making Europe. MM acknowledges the research programme Athena with project number 184.034.002, which is (partly) financed by the Dutch Research Council (NWO).    FMV acknowledges support from the European Union’s Horizon Europe research and innovation programme with the Marie Sk\l{}odowska-Curie grant agreement No. 101149685.  
        We thank Santiago del Palacio for valuable discussion and comments. 
        
\end{acknowledgements}

\section*{Data Availability}
The data underlying this article is publicly available in: {\em Pan-STARRS} \url{https://archive.stsci.edu/panstarrs/}and {\em AAVSO} \url{https://www.aavso.org}. Remaining data will be shared on reasonable request to the corresponding author.

   \bibliographystyle{aa} 

\begin{appendix}
\section{Observation log and light curve}
\begin{table*}[h!]
\centering
\caption{\label{table:data set} {\it XMM--Newton} EPIC-PN observation log of \srclong.}
    \begin{tabular}{lllll}
\hline \hline 
Epochs          &       ObsID           &               Date                                    &       State                                   &       Exposure time [ks]\tstrut\\
\hline
XMM1            &       0885000401      &               2024-02-26T14:20:50             &       high/soft                               &       61\tstrut\\[0.08cm]
XMM2            &       0885000701      &               2024-03-25T01:42:16             &       soft-to-hard transition      &       20\\ [0.08cm]
XMM3            &       0885000801      &               2024-03-27T12:14:55             &       soft-to-hard transition &    50\\ [0.08cm]
\hline
    \end{tabular}
\end{table*}

This dataset is part of a multiwavelength observational campaign with epochs defined by the optical observations from the {Very Large Telescope (VLT)}, performed with the {X-shooter spectrograph} \citep{2011A&A...536A.105V}; see Castro Segura et al. in preparation.\\

    \begin{figure}[h]
        \centering
        \includegraphics[width=1.1\linewidth]{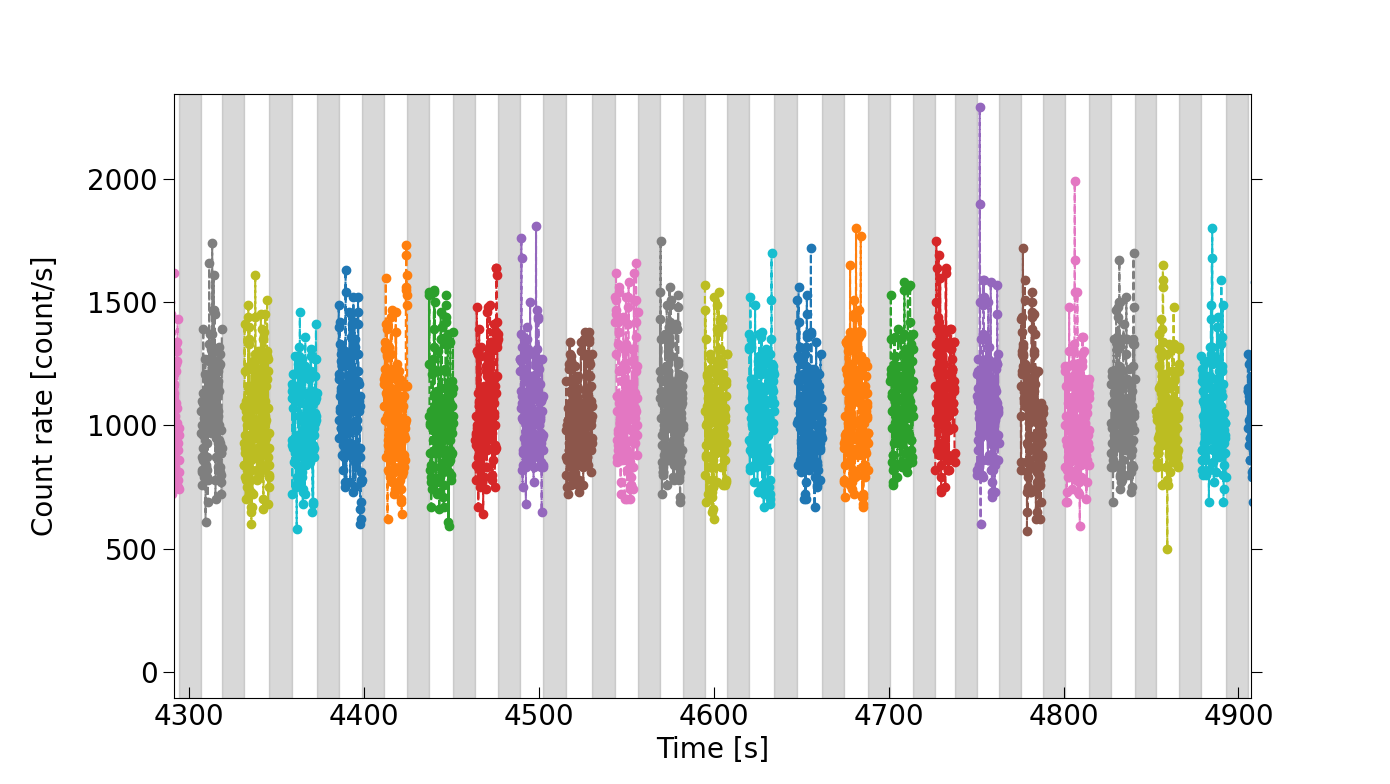}
        \caption{Zoomed XMM EPIC-PN light curves of \srclong\ in the 0.2--12 keV energy range, with a bin size of 0.1 s for the second observation (epoch XMM2). The shaded regions delimit the telemetry gaps and the colored data points are the good time intervals (GTIs).   }
        \label{fig: GTIs}
    \end{figure}

\newpage

\section{Hardness intensity diagram}
    \begin{figure}[!htb]
        \centering
        \includegraphics[width=1.1\linewidth]{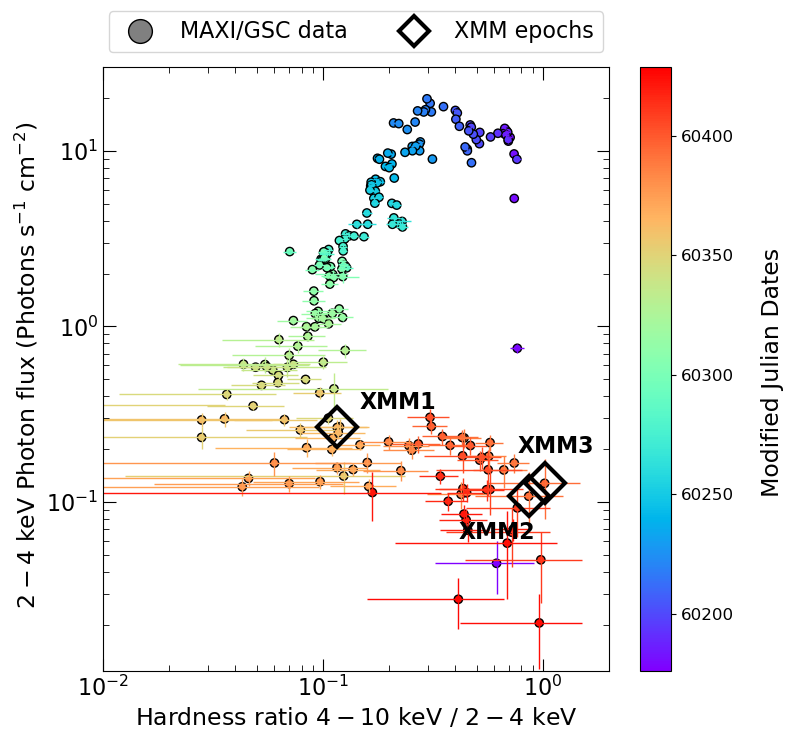}
        \caption{Hardness intensity diagram (HID) of \srclong\ from the 2023 outburst. The photon fluxes are derived from the MAXI/GSC instrument with a bin size of 1 day from the "on-demand process" archives. The hardness ratio is defined as the photon flux ratio in the 4--10 keV over the photon flux in the 2--4 keV energy band. }
        \label{fig:HID}
    \end{figure}

Figure \ref{fig:HID} shows the hardness intensity diagram (HID) of \srclong\ since the start of the 2023 outburst. The light curves were obtained from the Monitor of All-sky X-ray Image's \citep[MAXI;][]{2009PASJ...61..999M} Gas Slit Camera \citep[GSC; 2--30 keV;][]{2002SPIE.4497..173M, 2011PASJ...63S.635S, 2011GSC_PASJ...63S.623M}, in the 2--10 keV energy band, and the MAXI "on-demand process" archives\footnote{\url{http://maxi.riken.jp/mxondem}}. 
The position of the source in the HID during the XMM EPIC-PN observations are shown in black diamonds. The hardness ratio (HR) is defined as the ratio of the photon flux in the 4--10 keV energy band to that in the 2--4 keV energy band. \\

\end{appendix}

\end{document}